\documentclass[useAMS,usenatbib,fleqn]{mnras}
\usepackage{graphics,graphicx}
\usepackage{amsmath}
\usepackage{amssymb}
\usepackage{aas_macros}
\usepackage[usenames,dvipsnames]{xcolor}

\newcommand{\eagle}{{\sc eagle}}
\newcommand{\saga}{{\sc saga}}

\usepackage[para]{footmisc}

\title[]{Origins of carbon-enhanced metal-poor stars}

\author[]
{
Mahavir Sharma, Tom Theuns, Carlos S. Frenk and Ryan J. Cooke\\
Institute for Computational Cosmology, Department of Physics, Durham University, South Road, Durham, DH1 3LE, UK \\}

\begin{document}

\date{Submitted ---------- ; Accepted ----------; In original form ----------}


\maketitle

\begin{abstract}
We investigate the nature of carbon-enhanced metal poor (CEMP) stars in Milky Way (MW) analogues selected from the \eagle\ cosmological hydrodynamical simulation. The stellar evolution model in \eagle\ includes the physics of enrichment by asymptotic giant branch (AGB) stars, winds from massive stars, and type~Ia and type~II supernovae (SNe). In the simulation, star formation in young MW progenitors is bursty due to efficient stellar feedback, which enables poor metal mixing leading to the formation of CEMP stars with extreme abundance patterns. Two classes of CEMP stars emerge: those mostly enriched  by low-metallicity type~II SNe with low Fe yields that drive galactic outflows, and those mostly enriched by AGB stars when a gas-poor galaxy accretes pristine gas. The first class resembles CEMP-no stars with high [C/Fe] and with low [C/O], the second class resembles CEMP-s stars overabundant in s-process elements and high values of [C/O]. These two enrichment channels explain several trends seen in data: ({\em i}) the increase in the scatter and median of [C/O] at low and decreasing [O/H], ({\em ii}) the trend of stars with very low [Fe/H] or [C/H] to be of type CEMP-no, and ({\em iii}) the reduction in the scatter of [$\alpha$/Fe] with atomic number in metal  poor stars. In this interpretation CEMP-no stars were enriched by the stars that enabled galaxies to reionise the Universe.

\end{abstract}
\begin{keywords}
{stars: abundances -- nuclear reactions, nucleosynthesis, abundances -- dark ages, reionisation, first stars -- Galaxy: abundances, formation, halo	}
\end{keywords}

\section{Introduction}
Whereas hydrogen and most of the helium in the Universe were forged in the Big Bang \citep{Alpher48}, other elements were predominantly synthesised in stars \citep[e.g.][]{Burbidge57}. The standard model of stellar evolution includes three main channels of such \lq metal\rq\ enrichment: ({\em i}) Massive stars, $M\gtrapprox 6$~M$_\odot$, that burn He ashes hydrostatically yielding predominantly $\alpha$ elements (of which the nucleus consists of an integer number of $\alpha$ particles and hence an even number of protons, such as C, O, Mg, Ne, Si, etc), imprinting a characteristic \lq odd-even\rq\ elemental abundance pattern. These stars explode as core-collapse type II SNe, with neutron capture r-process and trans-Fe elements produced during explosive nucleosynthesis. ({\em ii}) Intermediate mass stars ($0.5\lessapprox M/{\rm M}_\odot<6$) that produce mainly C and O as
	well as s-process elements, that are brought to the surface and lost in a stellar wind or planetary
	nebula following the star's ascent up the asymptotic giant branch (AGB). ({\em iii}) Type Ia SNe with 
	significant Fe yields, and that are plausibly the result of mass transfer in a binary star that pushes the
	SN progenitor over the Chandrasekhar limit. The abundance pattern in stars reflects the extent to which
	these channels operate and how effective stellar ejecta mix with star-forming gas, see e.g. \cite{Nomoto13} for a review.

The lifetimes of the progenitors of these three channels are quite different. The short lifetimes ($\lessapprox 40$~Myr) of massive stars suggest that star forming gas will be rapidly enriched with $\alpha$ elements; the increase in the abundance of C from AGB stars, and Fe from type Ia SNe, delayed by $\sim 300$~Myrs. The consequences of such timed release of elements is evident in the high\footnote{The metallicity, $Z$, is the mass fraction in metals - elements more massive than helium. The common notation $[{\rm X}/{\rm Y}]\equiv \log(N_{\rm X}/N_{\rm Y})-\log((N_{\rm X}/N_{\rm Y})_\odot)$ denotes the number (or mass)   ratio of elements X and Y, relative to that in the Sun. Here we take the solar abundances from Table~1 of \cite{Wiersma09} but this may not be the case for the data to which we compare our results. Differences in the assumed solar abundances are small compared to the large variations discussed here.}
    $[\alpha/{\rm Fe}]$ abundances of elliptical galaxies in which star formation is rapidly suppressed
    following a burst \citep{Segers16}. On the other hand, if star formation is quiescent, abundances of stars in a low-$Z$ galaxy will increase slowly, with, for example, [C/O] increasing with metallicity $Z$, until eventually the abundances reflect the yield of the full stellar initial mass function (IMF).
    
The abundance patterns of Milky Way (MW) stars that formed relatively recently, such as the Sun for example, do not show large variations relative to the solar abundance pattern. However, the same is not true for extremely metal poor (EMP) stars (those with [Fe/H]$<$-3 in what follows; see \citet{Beers05} for a review), which can have [C/Fe]$\approx $+1 (carbon-enhanced metal poor stars, or CEMP stars in what follows). This may be caused by inefficient metal mixing, with the abundance pattern reflecting the yield of a few or even a single enriching supernova \citep[e.g.][]{Heger16}, which makes such stars valuable relics in studies of galactic archaeology \cite[e.g][]{Frebel15}.
	Abundance patterns different from that of the Sun are also detected
 	in some damped Lyman-$\alpha$ systems (DLAs) of low metallicity
 	([Fe/H]$\lessapprox -3$, e.g. \citet{Cooke11}).
 		
Abundances at low $Z$ may therefore provide valuable clues to stellar yields at low $Z$, and to the efficiency of metal mixing during early star formation. However, the variety of patterns seen at low $Z$ is baffling. The subclass of CEMP stars alone exhibits examples that show enhancements due to both s-\emph{and/or} r-process elements {\bf(\lq 	s\rq\ for slow and \lq 	r\rq\ for rapid)}, and stars with relatively normal abundances of neutron capture elements\footnote{Extremely metal poor stars enhanced in s-process elements, r-process elements, both, or neither, are denoted as CEMP-s, CEMP-r, CEMP-rs, and CEMP-no, respectively.} \citep[e.g.][]{Beers05}. CEMP-s stars may result from mass transfer from an AGB companion \citep[e.g.][]{Aoki07,Masseron10}, and indeed the majority - but crucially not all - CEMP-s stars show radial velocity variations \citep{Lucatello05,Starkenburg14,Hansen16}. \cite{Komiya07} claim that CEMP-r stars result from binary evolution as well, whereas for example \citet{Hansen15} suggest that the enrichment already happened in the star's birth-cloud. Below [Fe/H]=-3, most stars are CEMP-no, possibly because the natal gas was enriched by a single, early SN with high [C/Fe], as argued by e.g. \cite{Frebel06}.

Some observed abundance ratios at low metallicity do not appear to be consistent with any of the standard enrichment channels discussed above, possible pointing to the need for more exotic stellar models. Several authors have argued that the increase in [C/Fe] at low Z may be due to enrichment by population~III stars \cite[e.g.][]{Umeda03,Ryan05,Cooke14,Ishigaki14}	which might also explain the elevated [C/O] \citep{Akerman04,Fabbian09,Pettini14}, because their C yields are thought to be high \citep{Chieffi04}, see also \cite{Heger10,Limongi15}.  If confirmed, this would open up the exciting possibility of studying the nature of Pop.~III stars. Other abundance ratios in CEMP stars may be anomalous as well, including values of [O/Fe]$>$4 or [Ba/Fe]$<$-2 (see for example the review by \citealt{Frebel15}). \cite{Aoki14} see evidence for pair-instability SNe (PISN) in the abundance pattern of an EMP star. The abundance patterns in DLAs and their relation to early star formation, are discussed by \cite{Pettini08, Cooke11}. 

This variety of scenarios invoked to explain the abundance patterns of EMP stars and DLAs in the recent literature motivated us to examine the abundance patterns of stars in the \eagle\ suite of cosmological hydrodynamical simulations - which does not include any of these more exotic mechanisms. As discussed in detail in the next section, \eagle\ incorporates nucleosynthesis and enrichment from three channels: metallicity dependent nucleosynthesis in AGB stars, type Ia SNe, and elements produced during hydrostatic burning in massive stars and explosive nucleosynthesis in their subsequent type II SNe (but no specific Pop.~III yields nor PISN). We examine the resulting trends in abundances, and study how these patterns reflect both metal yields at low $Z$ and poor metal mixing during the onset of galaxy formation in the progenitors of galaxies like the Milky Way, the latter selection allowing comparison to data.

In Section 2 we describe the details of our simulation. In section 3 we describe
	the origin of CEMP stars in \eagle, including a discussion of why the simulation
	predicts the appearance of two branches, CEMP-s and CEMP-no. We compare to data
	when possible, and in particular in section~4. We summarise in section~5.

\section{The \eagle\ simulations}
\eagle\ (\citet{Schaye15}, hereafter S15) is a suite of cosmological hydrodynamical simulations based on the $\Lambda$ cold dark matter model of structure formation with parameters taken from the \cite{Planck14} paper. The simulations were performed with the {\sc gadget-3} code, based on the public Tree-SPH code of \cite{Springel05}, with changes to the numerical hydrodynamics scheme and new subgrid prescriptions for numerically unresolved physical processes relevant to galaxy formation (S15). The numerical parameters of the subgrid modules were calibrated to reproduce the redshift  $z\approx 0.1$ galaxy stellar mass function, galaxy sizes, and the stellar mass - black hole mass relation, as described by \cite{Crain15}. In this paper we use the simulation labelled L0100N1504 in Table~2 of \cite{Schaye15}; the SPH particle mass is $1.81\times 10^6$~M$_\odot$.

Full details of the subgrid modules and modification to the {\sc gadget-3} code used in \eagle\ can be found in S15, we summarise them very briefly here. Modifications to the code include the {\sc anarachy} SPH implementation described by Dalla Vecchia ({\em in prep.}, summarised by \citet{Schaller15}) and the time-step limiter of \cite{Durier12}. Cooling and photo-heating of cosmic gas in the presence of a pervasive and time-evolving UV/X-ray and cosmic microwave background is implemented as described by \cite{Wiersma09a}. Star formation is implemented following \cite{Schaye08}, whereby star forming gas particles are converted stochastically to collisionless \lq star\rq\ particles in a way that simulated galaxies follow the Kennicutt-Schmidt law \citep{Kennicutt98}.

Star particles in the simulation represent a simple stellar population (SSP) with a \cite{Chabrier03} stellar initial mass function (IMF) in the mass range [0.1,100]~M$_\odot$. As stars evolve, they enrich surrounding gas particles, spreading mass lost according to the SPH formalism; feedback from stars heat the gas as described by \cite{DallaVecchia12}. \eagle\ tracks 11 elements (H, He, C, N, O, Ne, S, Ca, Si, S, Fe) as well as a \lq total\rq\ metallicity variable, through the timed release of elements from the three channels summarised in the Introduction: SNe of types Ia and II (and winds from the massive star progenitors of type~II SNe) and AGB stars. Metallicity-dependent yields were taken from \citet{Woosley95} for type II SNe, from \cite{Marigo01} and \cite{Portinari98} for intermediate mass stars, and from model W3 of \cite{Thielemann03} for type~Ia SNe (see \citealt{Wiersma09} for full details and the Appendix for an illustration of some characteristic yields of these channels)\footnote{There can be an uncertainty in the yields, roughly of a factor of 2, however, this does not affect our results significantly as the abundances shown in the forthcoming sections vary by orders of magnitude. }. We track the contribution to Fe from SNe of type Ia, and the total mass from each of the three enrichment channels, separately. This allows us to determine which of the three nucleosynthetic channels tracked (AGB, Type~Ia, and massive stars and their SN type~II descendants) dominates the enrichment of a given element. To do so we combine the mass acquired through a particular channel, together with the nucleosynthetic yield of a particular element. For example, a high abundance of carbon relative to say oxygen would indicate the dominance of the AGB channel. This prediction can be verified by calculating the relative amount of metals acquired through the AGB channel compared to that acquired from massive stars.

Observed CEMP stars are often classified as CEMP-s or CEMP-no based on their barium abundance. Unfortunately the \eagle\ enrichment model does not track barium directly. However, since \eagle\ tracks the fraction of mass of any gas or star particle that it received from the AGB enrichment channel, we can estimate the Ba abundance approximately in post-processing. We do so by multiplying this mass fraction by the Ba yield of an AGB star. For the latter, we use the yields of a 3~M$_\odot$ AGB star from the models of \cite{Straniero14}.  This allows us to compare to observed abundances in section~\ref{sect:CEMP} below.

 The three enrichment channels discussed in the introduction are associated with stars of different initial mass (intermediate mass stars give rise to AGB enrichment, binary stars with low or intermediate mass components to type Ia SNe, and massive stars to type II SNe). If metal mixing is poor in real galaxies, as we will argue below, then it would be possible for mostly pristine star forming gas to be enriched predominantly by just one of these three channels - and hence for the appearance of stars that reflect mostly one of the three channels. However this is not possible in \eagle, because a simulation star particle represents a simple stellar {\em population}: as a simulation star particle ages it 	enriches its surroundings with both AGB and type Ia ejecta together, following a more rapid
	enrichment by massive stars. To allow us to study the enrichment by AGB stars
	separately\footnote{ This simulation limitation is not so crucial for enrichment by
	type II SNe, because the time-scale for enrichment by type II SNe is much shorter
	and gas can move significant distances between instances of type~II enrichment and that by the other two channels.}, we neglect enrichment by the type Ia channel altogether when computing stellar abundances. We can do so, because \eagle\ tracks the mass acquired through
	the type Ia channel separately. This implies that we underestimate the Fe
	abundances of stars.

To identify halos and galaxies in the simulation, we use {\sc subfind} \citep{Springel01, Dolag09} as described by \cite{McAlpine16}; the Milky Way-like galaxies whose stellar abundances we compare to observations below, are taken to be $z=0$ central galaxies that inhabit dark matter halos of mass $10^{12}\,{\rm M}_\odot< M_h <3\times 10^{12}{\rm M}_\odot$. We will refer to stars in Milky Way-like \eagle\ galaxies as \lq \eagle\ stars\rq\ in what follows.

To mitigate numerical sampling issues related to enrichment, \eagle\ additionally tracks \lq SPH smoothed\rq\ (as opposed to \lq particle\rq) abundances, as discussed by \cite{Wiersma09}. However, our confidence in the accuracy of predicted SPH-smoothed absolute abundances is still limited: we are confident that an \eagle\ star with smoothed Fe abundance of $[{\rm Fe/H}]<-2$, say, has indeed a low metallicity, but we cannot reliably distinguish stars with $-5<[{\rm Fe/H}]<-4$ from those with $-4<[{\rm Fe/H}]<-3$. However, relative abundances are not affected by sampling since they are sourced by the same star particles for all enrichment channels, and hence are much more reliable. To select candidate CEMP stars in \eagle, we will therefore select star particles with low abundance, [Fe/H]$<-2$, and examine their abundance pattern.

Star formation in low-mass halos, $M_h\lesssim 10^{10}{\rm M}_\odot$, is bursty in \eagle: the SFR is high when low-mass halos are gas rich, but the feedback from young massive stars may then remove a large fraction of the gas, dramatically reducing the SFR. We will refer to this phenomenon, whereby the gas fraction varies significantly as a function of time, as \lq breathing\rq. Although the level of stochasticity of the gas fractions in such small halos may be affected by numerical sampling of the feedback events, simulations of high-$z$ dwarf galaxies at much higher resolution typically show similar bursty behaviour \cite[e.g.][]{Wise14,Kimm14,Baldry16}, and therefore appear to be a generic prediction of current models. Such bursts may play an important role in reionisation \citep{Sharma16a, Sharma16b}.

\section{Two paths to CEMPs}

 In this section, we show that \eagle\ galaxies harbour two different branches of CEMP stars, which result from the interplay between the two nuncleosynthetic channels and strong feedback from type~II SNe in the first galaxies.

\subsection{An AGB origin for the [C/O] upturn at low $Z$}
\begin{figure*}
\includegraphics[width=0.8\linewidth]{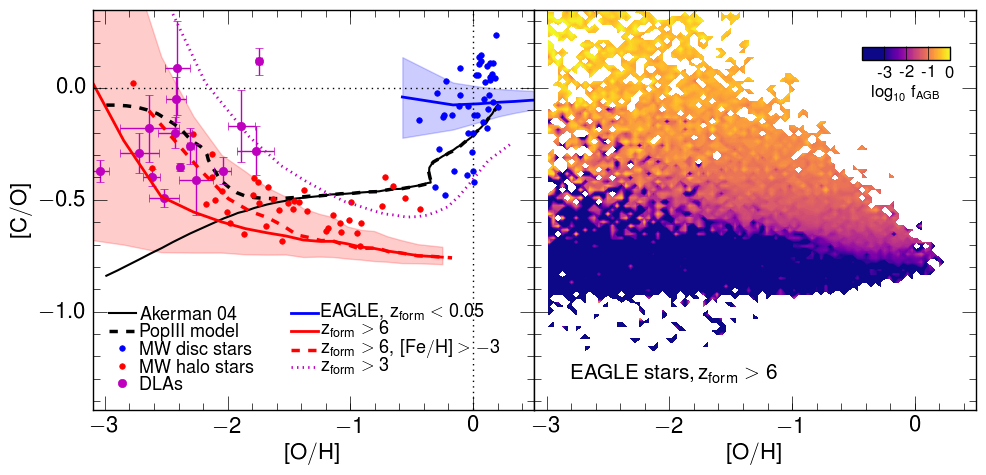}
\caption{\textit{Left panel:} [C/O] abundance as a function of [O/H]. \textit{Blue} and \textit{solid red} curves are the median relation for \eagle\ star particles that formed after $z=0.05$, or before $z=6$, respectively; the shaded blue and red regions enclose the 25$^{\rm th}$-75$^{\rm th}$ percentiles.
	The \textit{dashed red} curve is the median \eagle\ relation for stars formed before $z=6$ with [Fe/H]$>-3$. The dotted curve is the median \eagle\ relation for stars formed before $z=3$. The \textit{black curves} are population synthesis models from \citet{Akerman04}, without (solid) and with (dashed) a contribution from Pop.~III stars. \textit{Blue} and \textit{red} points are observed Milky Way disc and halo stars, respectively, taken from \citet{Fabbian09}. \textit{Magenta points with error bars} are the Damped Lyman Alpha (DLA) systems from \citet{Cooke17}. \textit{Right panel:} The median value of fraction, $f_{\rm AGB}$, from Eq.~\ref{eq:fAGB}, of enriched mass that originates from the AGB channel for the stars having a given [C/O] and [O/H].}
\label{fig:CO}
\end{figure*}

Standard enrichment models predict that [C/O] increases with metallicity, as illustrated by the model of \cite{Akerman04} plotted in black in Fig.~\ref{fig:CO}, because massive stars drive stronger winds at higher $Z$ \citep{Henry00,Carigi00,Akerman04,Cescutti09,Romano10}.  Consistent with this prediction, we find that \eagle\  stars formed recently, below $z=0.05$, and with metallicity [O/H]$\sim 0$, have [C/O]$\sim 0$, whereas older stars of lower metallicity, [O/H]$\sim -1$, have [C/O]$\sim -0.5$.
	
However, abundances of MW-stars display a surprising {\em upturn} in [C/O] below [O/H]$\sim -1$ \citep{Akerman04,Fabbian09}, and similarly high values of [C/O] are detected in low-$Z$ DLAs \citep{Pettini08,Cooke11b,Pettini14} and \cite{Cooke17} (red and magenta circles with error bars refer to MW stars and DLAs in Fig.~\ref{fig:CO}, respectively). It has been suggested that this upturn is a signature of enrichment by Pop.~III stars \citep{Akerman04, Pettini14}. In addition to an upturn, the {\em scatter}\footnote{Error bars on the observed abundances of stars are not plotted in the figure, but are typically small compared to the scatter between points.} in [C/O] increases dramatically with decreasing [O/H].

Abundances in \eagle\ (red curve) show a similar upturn and increase in scatter (shaded red region), even though Pop.~III stars are not part of the model. Because the simulation tracks enrichment by each channel separately, we know that the high values of [C/O] at low [O/H] reflect enrichment by AGB stars instead. This is demonstrated in the right panel of Fig.~\ref{fig:CO}, where we plot [O/H] versus [C/O] for \eagle\ stars formed early, before $z=6$, coloured by the ratio 
\begin{equation}
f_{\rm AGB}\equiv {m_{\rm AGB}\over m_{\rm AGB} + m_{\rm SNII}}\,,
\label{eq:fAGB}
\end{equation}
   where $m_{\rm AGB}$ and $m_{\rm SNII}$ are the metal mass received by the precursor gas particle of a star from the AGB and massive stars channels, respectively. Stars with $f_{\rm AGB}=1$ are {\em only} enriched by AGB stars (coloured yellow), and have the most extreme values of [C/O]$\gtrsim1$. 
   
The stars that have such low [O/H] and high [C/O] formed before $z=6$ (corresponding $\sim 1~$Gyr after the Big Bang, and only $\sim 700~$Myrs after the formation of the first stars in this simulation, at $z\sim 15$), which, maybe somewhat surprisingly, is late enough for the AGB channel to become active. Indeed, the stellar evolution models of \cite{Marigo01} and \cite{Portinari98} used in \eagle, already yield significant AGB enrichment 300~Myrs after the stars formed. Since the upturn is due to carbon produced by AGB stars in \eagle, oxygen - which is not synthesised significantly in AGB stars (yields of [C/O]$>1$, see below)- is {\em low}. We recall that type~II SNe  and their massive progenitor stars, produce both carbon and oxygen, and their yields do not become highly super-solar in [C/O]. Therefore, if our interpretation is correct, the high [C/O] stars are not so much carbon-enhanced as oxygen-poor. Somehow, the birth cloud of these stars avoided being enriched by massive stars for long enough, $\gtrapprox 300$ Myr, to allow AGB progenitors to evolve and release carbon.

The model by \cite{Salvadori12} predicts a population of [C/O]$\gtrapprox 0.5$ systems with [O/H]$\lessapprox -2$, which results from enrichment of primordial gas by Pop.~III stars. They associate the 	gentle increase in [C/O] of those systems with increasing [O/H] with enrichment by low-$Z$ AGB stars. In contrast, the origin of the high values of [C/O] at low [O/H] in \eagle\ is not due to Pop.~III enrichment, since the model does not include such stars, and we explore the nature of this enrichment in more detail next.

\subsection{Breathing and poor metal mixing}
\begin{figure}
\includegraphics[width=\linewidth]{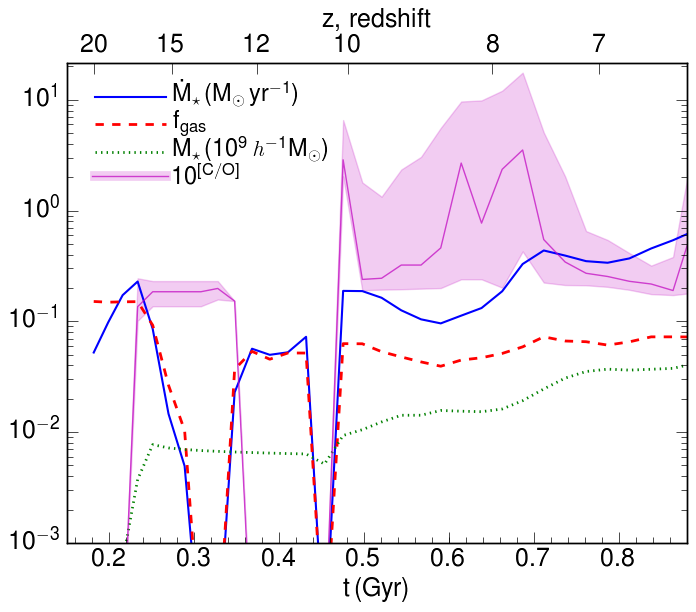}
\caption{Evolution of the star formation rate ({\em solid blue line}), gas fraction ({\em dashed red line}) and the stellar mass ({\em dotted green line}) for a typical \eagle\ Milky Way progenitor as a function of cosmological time.  The thin magenta line shows the median of carbon to oxygen abundance ratios of star particles formed within the last $100$ million years of a given redshift, with the shaded region showing the 20th and 80th percentile.  Star formation at such low-mass in \eagle\ is very bursty, with periods of high $\dot M_\star/M_\star$ associated with strong outflows, followed by quiescent periods in which the galaxy is gas poor.}
\label{fig:bursty}
\end{figure}

Enrichment due to massive short-lived stars should occur soon after a galaxy starts forming stars. As a consequence, the second generation of stars would be expected to show the signature of such  stars - high abundances of $\alpha$ elements, for example. However, the abundance pattern of newly formed stars will only reflect the average yields of their precursors if the ejecta of stars mix well with star-forming gas. In \eagle,  as shown in Fig.~\ref{fig:bursty}, and in many other simulations as well \citep[e.g.][]{Wise14,Kimm14,Baldry16}, star formation is very bursty and the gas content varies significantly with time in the low-mass progenitors in which low $Z$ MW-stars form - we refer to this as breathing modes. When the galaxy exhales by ejecting gas, star formation mostly shuts down until new, predominantly pristine, gas accretes. If this happens on a timescale of the order of 300~Myrs later, then accreted pristine gas may be enriched by AGB stars, and stars that
	form from this gas may show the signature in their abundance pattern of AGB yields - high [C/O], for example. To demonstrate that this scenario occurs in \eagle, we first show that stars that
	form in low mass \eagle\ galaxies  show very large variations in the value of 
	$r\equiv f_{\rm AGB}/f_{\rm SNII}$, predicting low $Z$ signatures of pure AGB/type II
	enrichment. We next demonstrate that this is related to the gas fraction of the galaxy at the
	time the stars form.

\begin{figure}
\includegraphics[width=\linewidth]{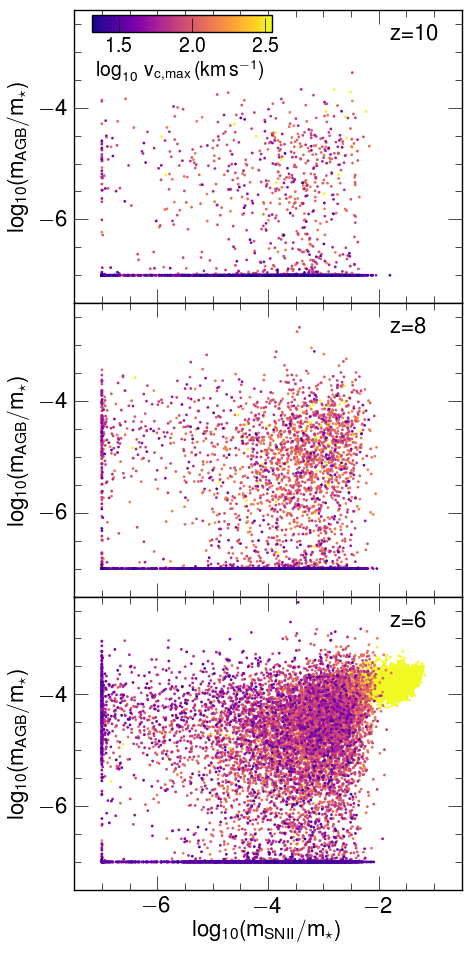}
\caption{The fraction of a star particle's mass acquired from the AGB enrichment channel, $m_{\rm AGB}/m_\star$,  as a function of the fraction of mass acquired from the SNe type II channel, $m_{\rm SNII}/m_\star$, for stars born 20~Myr before redshift $z=6$, 8 and 10 (bottom to top panels, respectively). Star particles with $m_{\rm AGB}/m_\star<10^{-7}$ are plotted at $10^{-7}$, and similarly for $m_{\rm SNII}/m_\star$. Star particles are coloured according to the maximum circular velocity, $v_{\rm c, max}$, of their parent halo, as indicated in the colour bar.}
\label{fig:mix}
\end{figure}

\begin{figure}
\includegraphics[width=\linewidth]{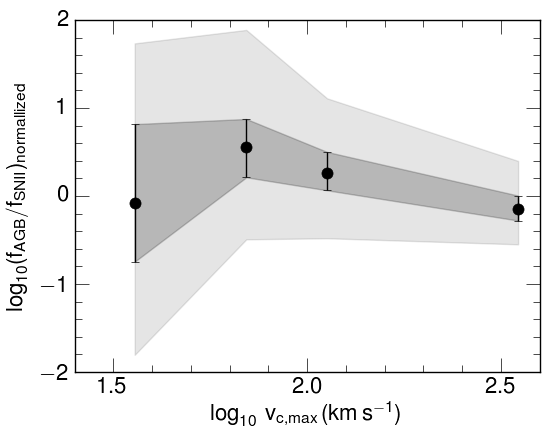}
\caption{Ratio $f_{\rm AGB}/f_{\rm SNII}$  of \eagle\ star particles that formed before $z=6$, normalised to the median value of stars that form in haloes with $v_{\rm c,max}>100$ km~s$^{-1}$, as a function of the maximum circular velocity of their parent halo. Symbols show the median ratio, with error bars and dark grey shading enclosing the 40$^{\rm th}$-60$^{\rm th}$ percentiles, light grey shading encloses the 20$^{\rm th}$-80$^{\rm th}$ percentiles. The large increase in the scatter of $f_{\rm AGB}/f_{\rm SNII}$ towards lower $v_{\rm c, max}$ demonstrates that stars that form in such small halos are increasingly enriched either exclusively by AGB, or exclusively by type II SNe.}
\label{fig:vcmix}
\end{figure}
\begin{figure}
\includegraphics[width=\linewidth]{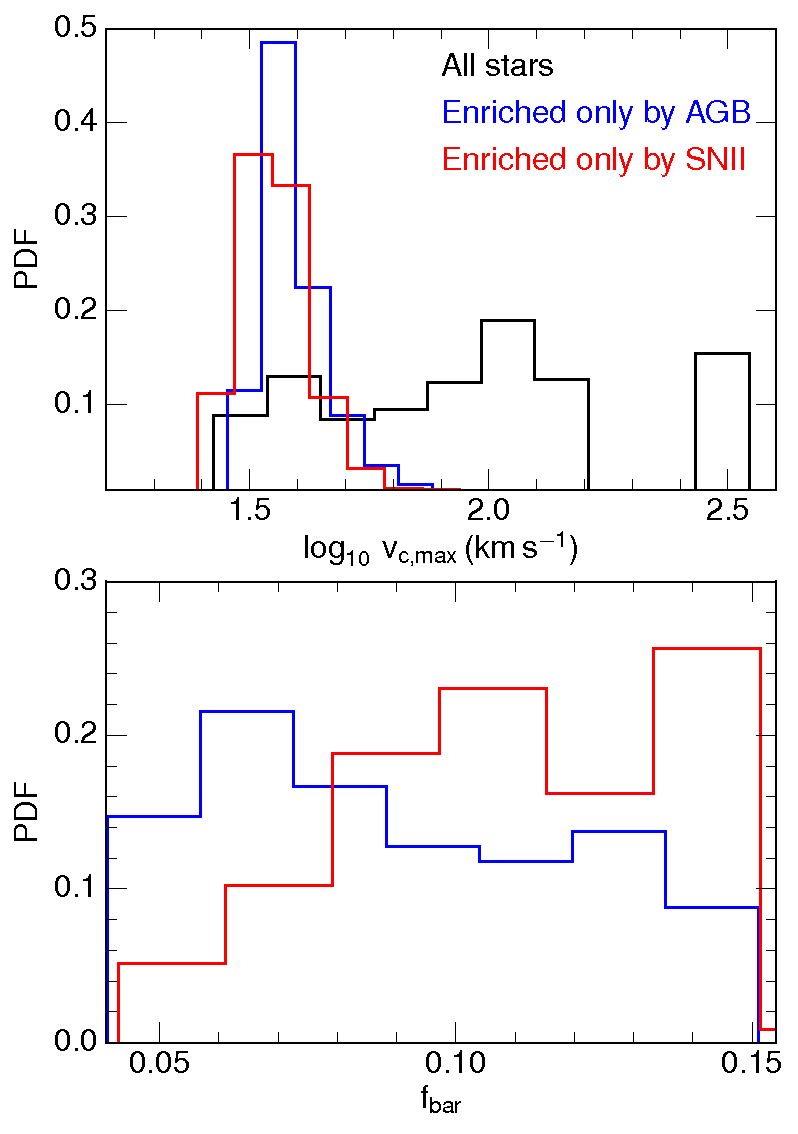}
\caption{\textit{Upper panel:} Fraction of \eagle\ stars formed in halos of a given maximum circular velocity, $v_{\rm c,max}$, 20~Myr before $z=6$. \textit{Black}, \textit{blue} and \textit{red} histograms refer to all stars, stars enriched only by AGB, and stars enriched only by massive stars, respectively. Each histogram is separately normalised to unit integral. \textit{Bottom panel:} baryon fraction of \eagle\ stars that form in halos with $v_{\rm c,max}\leq 50$~km~s$^{-1}$, within 20~Myr before $z=6$. \textit{Blue} and \textit{red} histograms show stars enriched only by AGB, and stars enriched only by massive stars, respectively. Each histogram is normalised to unit integral.}
\label{fig:mix_hist}
\end{figure}

Figures \ref{fig:mix} and \ref{fig:vcmix} demonstrate our first claim. Galaxies in halos at $z\gtrapprox 6$ with low maximum circular velocity, $v_{\rm c, max}\lessapprox 50$~km~s$^{-1}$, form stars with a wide range of $r=f_{\rm AGB}/f_{\rm SNII}$. At $z=10$ (top panel of Fig.~\ref{fig:mix}), they form stars with the abundance patterns characteristic for massive star enrichers (plotted at $f_{\rm AGB}=10^{-7}$) but few show evidence for AGB enrichment, simply because the Universe is too young ($\sim 0.5$~Gyr) for the AGB channel to operate actively. At later times (middle and bottom panels at $z=8$ and $z=6$ in Fig.~\ref{fig:mix}), more stars with high $f_{\rm AGB}$ appear, including stars with $f_{\rm AGB}\approx 1$ which exhibit nearly pure AGB abundances (e.g. [C/O]$>1$). Such extreme abundance patterns occur far less at higher $v_{\rm c, max}$; see for example the abundant cloud of orange/yellow points ($v_{\rm c, max}>100$km~s$^{-1}$), which correspond to stars that form in galaxies in which AGB and type~II channels are well mixed. Figure~\ref{fig:vcmix} shows in more detail that the scatter in $r$ increases dramatically below $v_{\rm c, max}\sim 50$~km~s$^{-1}$, consistent with the first claim.

The second claim is demonstrated in Fig.\ref{fig:mix_hist}: stars with extreme AGB/type~II abundances form predominantly in halos with $v_{\rm c, max}\lessapprox 50$~km~s$^{-1}$ (top panel). Those stars predominantly enriched by AGB stars form in halos that, in addition, have low baryon fractions (bottom panel). The latter are small halos where gas has been removed by a previous star burst, with pristine cosmologically accreted gas now being enriched by the early generation of AGB stars, imprinting their characteristic AGB pattern on any newly forming stars. Conversely, gas rich dwarfs (with high baryon fractions) form stars that may have the characteristic pattern of yields for massive stars. The associated type~II SNe then power the outflow that causes the galaxy to become gas poor.

How realistic is it that this scenario also applies to early galaxies, given the limitations of \eagle? It is based on two aspects of the simulation: ({\em i}) large variations in the gas fraction of dwarfs (breathing), and ({\em ii}) poor metal mixing of stellar ejecta. A large mass loading factor - the ratio $\beta=\dot M_{\rm wind}/\dot M_\star$, of the galactic outflow rate to the star formation rate - for high-$z$ dwarfs, appears to be an essential ingredient of simulations \cite[e.g.][]{Muratov15}. Although galactic winds are indeed ubiquitously observed at high-$z$, measuring $\beta$ is challenging; see for example the review by \cite{Veilleux05}. However, metals are observed in the intergalactic medium \citep{Cowie95}, even at low density \cite[e.g.][]{Schaye03}, and it is likely that these were deposited there by galactic outflows originating predominantly from dwarf galaxies \citep{Madau01,Theuns02, Booth12}. Such wind episodes may also explain the high escape fractions of ionising photons, needed to reionise the Universe \citep{Sharma16a}. Given this evidence we posit that this aspect of our model is relatively well established.
	
How about the poor metal mixing of enriched gas? If, as is likely, winds are (at least partially) powered by massive stars, then it would not be surprising if the metallicity of the wind were {\em higher} than that of the general ISM, since it is the hot ejecta that provides the buoyancy for the gas to escape. This is indeed seen in the parsec-resolution wind simulations of \cite{Creasey15}. Therefore, it is plausible that dwarfs can indeed lose a significant fraction of SNe type II products in a galactic outflow. But is it then possible for the remaining gas to be enriched by  massive stars to the extremely low levels seen in the simulation? Observations by \cite{James16} of nearby star forming galaxies show that metals are poorly mixed on scales of $\sim$~50 parsecs, which they attribute to poor metal mixing around young, star forming regions.  Another line of observational evidence comes from the work of \cite{Schaye07}, who identified a large population of photo-ionised, compact (sizes 100~pc), metal rich ($Z\sim Z_\odot$) clouds in the redshift $z\sim 2$ intergalactic medium.  They attribute the existence of these clouds to poor metal mixing. Given these two lines of observational evidence, we argue that poor 	metal mixing in the $z\gtrapprox 6$ dwarfs is at least plausible. 
	
 The scenario of poor metal mixing in breathing galaxies makes another testable prediction: if metal mixing is indeed poor, especially during the early stages of star formation in a galaxy, we would expect to see stars enriched (almost exclusively) by SNe of type II as well. We investigate observational evidence for this next.

\subsection{The origin of stars with high [C/Fe] at {\em low} [C/O]}

\begin{figure}
\includegraphics[width=\linewidth]{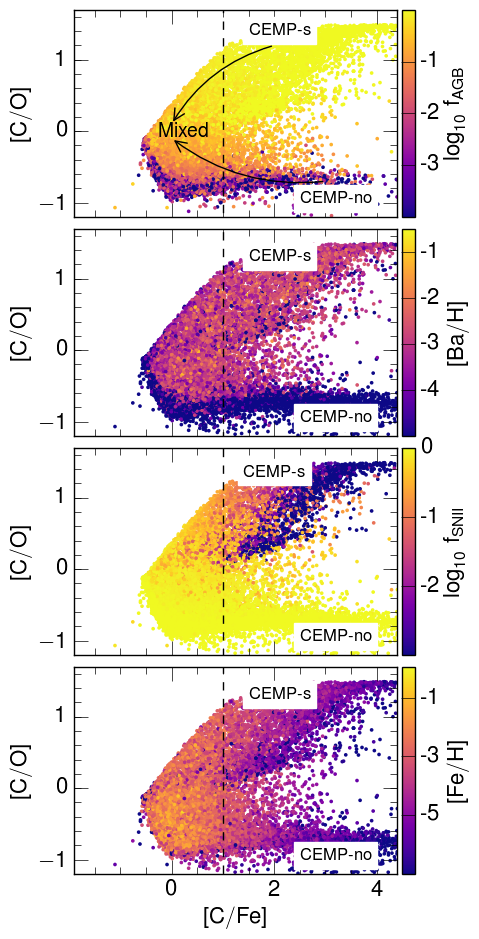}
\caption{[C/O] versus [C/Fe] in stars that formed before $z=6$ in \eagle. From top to bottom, star particles are colour coded according to their AGB fraction ($\log(f_{\rm AGB})$), barium to hydrogen ratio ([Ba/H]), SNe type~II fraction ($\log(f_{\rm SNII})$) and [Fe/H] abundance, respectively.  Stars with high $f_{\rm AGB}$ have high [C/O]$>0.5$, high [Ba/H] ($>-2$) at high [C/Fe] (yellow points in the top two panels), and conversely stars with high $f_{\rm SNII}$ (and hence low $f_{\rm AGB}$) have low [C/O]$<0$ at high [C/Fe] (yellow points in the third panel from the top). There are therefore two classes of CEMP stars in \eagle. Stars with low iron abundance, [Fe/H]$<-3$ (dark colours in the bottom panel) can be of either type, although those with the most extreme values of [Fe/H] ($<-5$) appear on the bottom branch only: these have [C/O]$<-0.5$. The \textit{arrows} in the top panel indicate the general increase with time of metallicity and amount of mixing of nucleo-synthetic channels.}
\label{fig:two}
\end{figure}
\begin{figure}
\includegraphics[width=\linewidth]{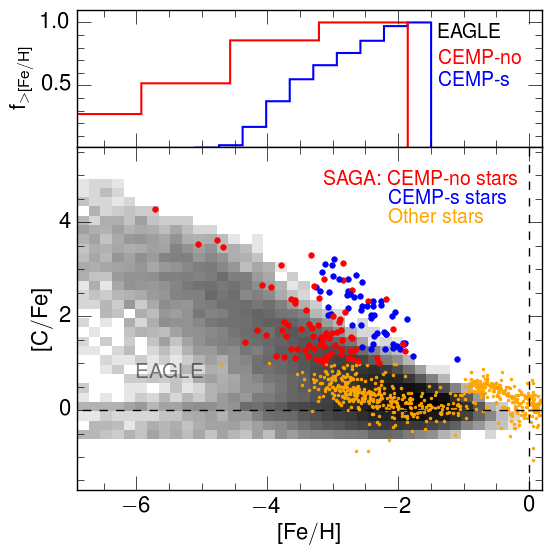}
\caption{{\it Bottom panel}: [C/Fe] as a function of [Fe/H].  The distribution of  \eagle\ stars formed before $z=6$ is shown using a grayscale 2D histogram. Observed stars, taken from the \saga\ database, are shown as red and blue circles for CEMP-no ([C/Fe]$>1$, [Ba/H]$<-2$) and CEMP-s stars ([C/Fe]$>1$, [Ba/H]$>-2$), respectively, with the remainder plotted as orange dots. Cumulative probability distribution of [Fe/H] for \eagle\ CEMP-no and CEMP-s stars is shown as red and blue histograms, respectively.
} 
\label{fig:CFe}
\end{figure}
\begin{figure}
\includegraphics[width=\linewidth]{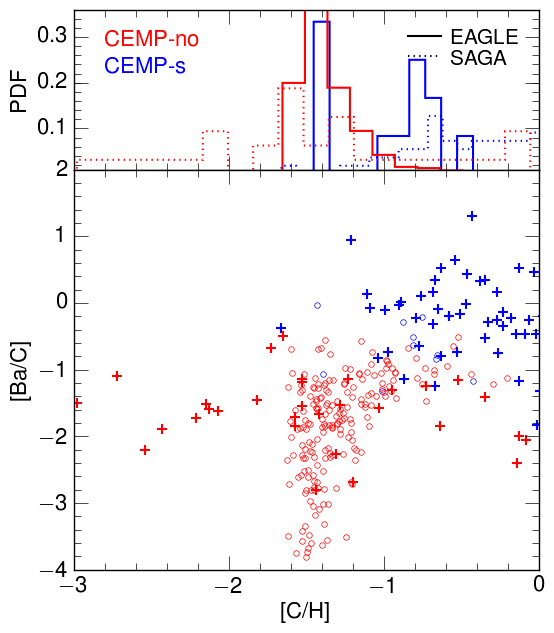}
\caption{\textit{Bottom panel:} [Fe/H] as a function of [C/H] for CEMP stars ([C/Fe]$>1$). \eagle\ stars classified as CEMP-no (CEMP-s) are shown as \textit{red} (\textit{blue}) circles, the text describes how these are selected. Stars from the \saga\ data base \citep{Suda08} are shown as plus symbols using the same colour scheme, and classified as described in Fig.~\ref{fig:CFe}. \textit{Top panel:} \textit{Solid histograms:} distribution of [C/H] for the \eagle\ stars classified as CEMP-no (\textit{red}) and CEMP-s (\textit{blue}). \textit{Dashed histograms:} same but for observed stars taken from the \saga\ database. Each histogram is normalised separately to unit integral.}
 \label{fig:CH}
\end{figure}
The combination of poor metal mixing and the existence of two channels of carbon production (AGB and  massive stars) gives rise to two classes of CEMP stars in \eagle: those enriched by AGB stars (which produce s-process elements such as Ba but do not produce Fe), and those enriched by  massive stars and their type~II SNe descendants with Fe-poor ejecta. In the models of \cite{Woosley95} used in \eagle, the latter occur for a wide range of progenitor masses at low metallicity, $Z\lessapprox 0.004$, and also for progenitor masses of $M_\star\approx 30$~M$_\odot$ for $Z=0.02$ (see also Fig.~\ref{fig:yield} below).
		
Massive stars are the first to enrich their surroundings as a galaxy begins to form stars in \eagle\footnote{We remind the reader that the simulation does not include any Pop.~III stars}. Their yields are extremely high in [C/Fe] and slightly subsolar in [C/O] (see the Appendix). As time progresses and lower mass type~II SNe explode, the enrichment pattern shifts to yields with lower [C/Fe] and values of [C/O] still within $\sim 0.8$~dex from solar. The timescale for this initial enrichment is of course very short, with the $\sim 10~$Myr lifetime of a 20~M$_\odot$ star much shorter than the $\sim 300$~Myr delay of the appearance of the first AGB events. As time progresses, we therefore expect the abundance of star forming gas that is enriched by massive stars to shift along the bottom arrow in the top panel of Fig.~\ref{fig:two}.
	
However, if feedback from these massive stars is able to eject a significant fraction of the	 star-forming gas, then star formation may temporarily halt. When it resumes, following cosmological accretion of mostly pristine gas, AGB stars may enrich star-forming gas yielding stars with high [C/Fe] and [C/O]$>1$.  Such stars correspond to the upper branches at high [C/Fe] in Fig.~\ref{fig:two}. As time progresses and the potential well that hosts the galaxy grows in mass, dramatic outflows following bursts diminish, and AGB and type~II (and type~I as well) ejecta mix well, so that when [Fe/H]$\lessapprox -2$, the ratios of [C/Fe] and [C/O] of star forming gas eventually approach solar values.

This scenario predicts that at very low $Z$, the value of [C/O] should increase with decreasing [O/H] fastest for stars with [Fe/H]$\gtrapprox -3.5$. Indeed, the MW progenitors in which these stars form are sufficiently evolved to host AGB stars, and it is their C-rich but O-poor ejecta that cause [C/O] to increase. At even lower [Fe/H], the MW progenitor is too young to host significant numbers of AGB stars and enrichment is mostly by  massive stars: it is in these MW progenitors that the stars with lower [C/O] at very low [O/H] in Fig.~\ref{fig:CO} form. 

\section{Observational evidence for the two paths of CEMP formation}
\label{sect:CEMP}
In the previous section we demonstrated that there are two distinct paths that yield CEMP stars in \eagle\ - enrichment by AGB stars when the galaxy is gas poor, and enrichment by low Z type II SNe when the galaxy is gas rich. While both paths lead to stars with high [C/Fe] characteristic of CEMP stars, abundance ratios of other elements can be quite different, for example the AGB path leads to stars with much higher [C/O]. It should be possible to directly test the existence and frequency of occurrence of these different paths by measuring the fraction of CEMP stars with high s-process element abundances (CEMP-s stars).

The s-process in AGB stars (slow neutron capture when the neutron bombardment of seed nuclei occurs at a rate that is slow compared to the $\beta$ decay rate of neutron rich nuclei) is thought to  originate overabundances of Sr, Yr, Zr, and Ba, La, Ce, Pr and Nd, see e.g. \cite{Sneden08} for a review. Therefore the prediction from \eagle\ is that the high [C/O] CEMP stars are of sub-type CEMP-s, whereas the low [C/O] stars are not s-process enhanced and of type CEMP-no (see \cite{Beers05} for definition of these classes).

Because both Ba and Fe produce relatively strong lines in stellar spectra, even at low abundance,
	and with Ba characteristic for s-process nucleosynthesis, a high value of [Ba/Fe] is often used
	observationally to distinguish between CEMP-s ([Ba/Fe]$>1$) and CEMP-no classes. 
	Other abundance ratios have been used as well. The dominance of the AGB channel over that of type II
	SNe in \eagle, is quantified unambiguously by $f_{\rm AGB}$ from Eq.~(\ref{eq:fAGB}) - the fraction of
	metal mass received from the AGB - and hence this is an excellent measure of whether a star is of type
	CEMP-s (high $f_{\rm AGB}$) or CEMP-no (low $f_{\rm AGB}$), as we did in the previous section. However
	this ratio is of course not directly observable. The two branches of CEMP stars shown in
	Fig.~\ref{fig:two} both result from channels that may not produce any Fe at all (the AGB channel for
	the top branch, and low $Z$ type II SNe that also do not produce Fe). This, combined by our
	neglect of metals produced by type Ia SNe for the reasons explained in Section 2, makes [Ba/Fe]
	a poor classifier for CEMP-s stars in \eagle. Fortunately, Fig.~\ref{fig:two} demonstrates
	that $f_{\rm  AGB}$ correlates very well with [Ba/H] as well as with [Ba/C], and we will therefore
	classify a CEMP star as CEMP-s when [Ba/H]$>$-2. This classification applied to
	observed stars compiled in the {\sc saga} database\footnote{\url{http://saga.sci.hokudai.ac.jp/
	wikidoku.php}} by \cite{Suda08} yields the same division in CEMP-s versus CEMP-no as a classification
	based on [Ba/Fe]. In summary we select  CEMP stars in {\sc eagle} using the criteria [C/Fe]$>1$
	following \cite{Beers05} and denote them CEMP-s if [Ba/H]$>-2$ and CEMP-no otherwise.
	
\subsection{CEMP-s versus CEMP-no stars: comparison with observations}

In Fig.~\ref{fig:CFe}, we plot [C/Fe] versus [Fe/H] for \eagle\ stars (depicted as a 2D greyscale histogram) and compare this to abundance ratios of Milky Way stars taken from the \saga\ database. The observed stars are collected from a diverse set of observational surveys with a variety of selection criteria. The database is therefore not a complete sample of CEMP stars.

In red giant branch (RGB) stars, carbon can be burned to nitrogen \citep{Roederer14,Placco14}. When this happens, the measured (surface) carbon abundance does not reflect the initial carbon abundance. We therefore exclude observed RGB stars (using the criterium that their surface gravity $\log_{10}(g/({\rm cm}~{\rm s}^{-2}))<3.2$) from Figs.~\ref{fig:CFe} and \ref{fig:CH}. However, we do show RGB stars in Figs.~\ref{fig:COFe} and \ref{fig:X} below: their [C/$\alpha$] are lower limits, but [$\alpha$/Fe] is unaffected by nuclear burning.

In both \eagle\ and \saga, the scatter and the median value of [C/Fe] increase with decreasing [Fe/H] (Fig.~\ref{fig:CFe}). The observations show a cluster of points with [C/Fe]$\sim 2$ at $-3\lessapprox {\rm [Fe/H]}\lessapprox -2$, most of which are CEMP-s stars and a significant fraction of these are believed to be binary stars. The enhancement in carbon and s-process elements in these binary stars likely results from mass transfer. Since \eagle\ does not include binary star evolution, it comes as no surprise that they are absent from the simulation.

At low values of [Fe/H]$ \lessapprox -3$, observed stars with [C/Fe]$>1$ are mostly of type CEMP-no (see red dots in Fig.~\ref{fig:CFe}, although observed CEMP-s stars with [Fe/H]$<-3$ do exist, \cite{Yoon16}), suggesting a link between the level of Fe enrichment and CEMP-no nature \citep[e.g.][]{Aoki07, Yong13}. This is true in \eagle\ as well: most stars with low [Fe/H] are of type CEMP-no (see also the upper panel of Fig.~\ref{fig:CFe}). 
	
\citet{Yoon16} argue that [C/H] is tighter correlated with s-process enhancement than [Ba/Fe] or [Fe/H] discussed by \cite{Aoki07}, with CEMP-no stars dominating at low [C/H] and CEMP-s stars at higher [C/H]. This is true in \eagle\ as well: CEMP stars with [C/H]$>-1.5$ tend to be high in [Ba/C] and are s-process enriched (blue circles), whereas those with lower C abundance are mostly CEMP-no
 	(red circles), as illustrated in Fig.~\ref{fig:CH}. A similar division is apparent for observed stars
 	in the {\sc saga} database (blue and red crosses, respectively). The top panel of the figure depicts
 	the corresponding probability distributions: we remind the reader that the \saga\ database is not
 	complete nor unbiased.

Although the distribution of [C/H] and the correlation with s-process enhanced CEMP stars clearly differ in detail between \eagle\ and the observed stars from the {\sc saga} database, nevertheless we see that \eagle\ reproduces the observed trend of CEMP-no stars dominating at lower [C/H]. The underlying reason is that such stars formed out of mostly pristine gas enriched very early on by high mass, low $Z$ stars, whereas the CEMP-s stars appear later on, forming from gas enriched with C by AGB stars. Given the longer evolutionary timescale of AGB stars, this gas is generally more enriched including
 having a higher Fe abundance, as we discussed in reference to Fig.~\ref{fig:CFe}.

Summarising, we find that in \eagle, CEMP-s and CEMP-no stars form as a result of two very different enrichment paths. CEMP-s stars form due to AGB stars enriching primordial gas accreting onto a gas poor dwarf galaxy. They are iron poor because the AGB channel does not produce iron. CEMP-no stars, on the other hand, result from enrichment by metal poor massive stars, whose type~II iron yields are very low. These two branches standout when plotting [C/O] vs [C/Fe] at high [C/Fe], with the first branch having [C/O]$\gtrapprox 1$ and the second branch having [C/O]$\lessapprox -0.5$. This scenario also predicts correctly the observed predominance of CEMP-no stars at low [C/H]. We examine observational evidence for enrichment by low $Z$ SNe for the appearance of the lower branch of CEMP stars in
	Fig.~\ref{fig:two} next.

\subsection{Signatures of low-$Z$ enrichment by massive stars}
\begin{figure}
\includegraphics[width=\linewidth]{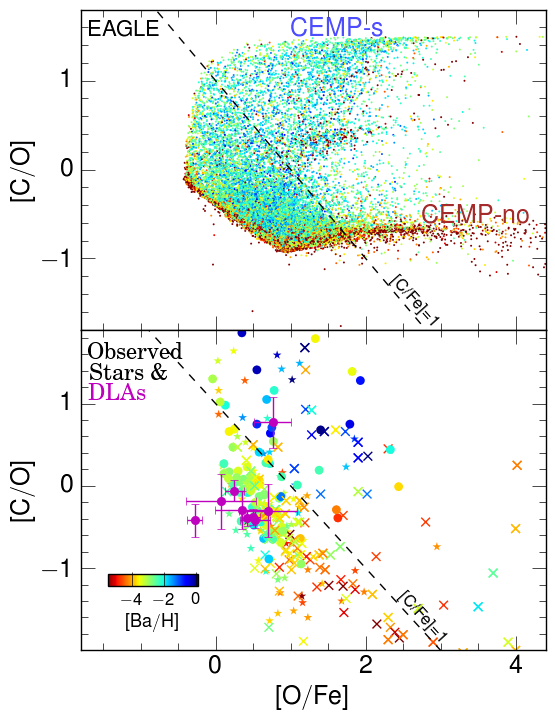}
\caption{[C/O] as a function of [O/Fe]. \textit{Top panel:} distribution of \eagle\ stars that formed before $z=6$ with \textit{cyan-green-red} colour map representing the variation in [Ba/H]. 
\textit{Bottom panel:} distribution of metal poor MW stars with [Fe/H]$<-2$, taken from the {\sc saga} database (\citealt{Suda08}) . Observed stars, coloured according to their [Ba/H], are shown as circles; crosses for limits. In this panel we also include red giants (surface gravity $\log_{10}(g/({\rm cm}~{\rm s}^{-2}))<3.2$); they are plotted using star symbols. The [C/O] of RGB stars is a lower limit as surface carbon may be depleted due to nuclear burning. Purple symbols with error bars are DLAs data from \citet{Cooke17}.}
\label{fig:COFe}
\end{figure}

\begin{figure*}
\includegraphics[width=0.8\linewidth]{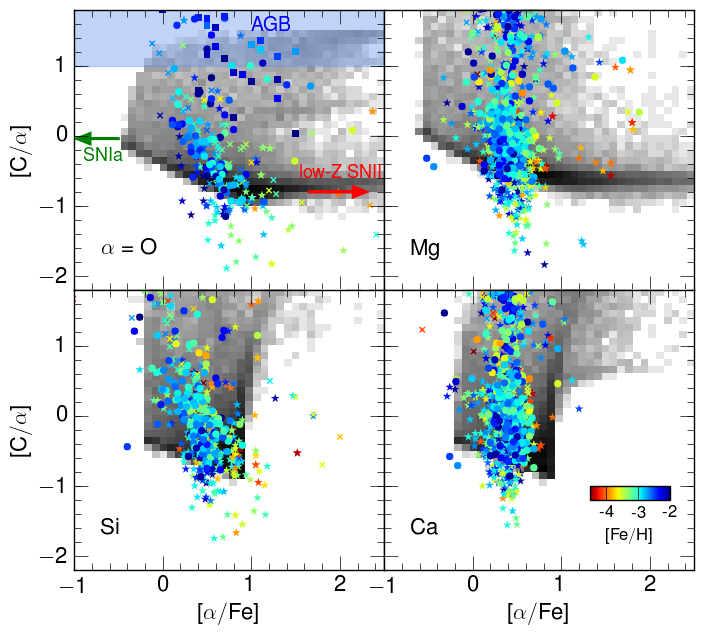}
\caption{The carbon to $\alpha$-element abundance ratio [C/$\alpha$] as a function of [$\alpha$/Fe], for $\alpha=$O, Mg, Si and Ca, respectively (as indicated in each panel). Results from \eagle\ are underlaid in \textit{grey}. Labels \lq AGB\rq\ (blue region), and \lq low-$Z$ SNII\rq\ indicate the approximate loci where the corresponding channels dominate enrichment in \eagle. The effect of including Fe enrichment from type Ia SNe is indicated by the green arrow, which is plotted at the [C/O] yield of type~Ia SNe (see the Appendix). Abundance ratios of Milky Way stars taken from the \saga\ database (\citealt{Suda08}) and \citet{Kennedy11} are shown as circles and squares respectively. The datapoints are coloured according to their [Fe/H] abundance (colour bar in bottom right panel). Circles and squares denote detections, crosses show limits in case of none detection of the 	$\alpha$ element. RGB stars are plotted with star symbols, see Fig.~\ref{fig:COFe}: their [C/$\alpha$] is a lower limit due to
	depletion of carbon through nuclear burning.}
\label{fig:X}
\end{figure*}

 The two classes of CEMP stars ([C/Fe]$>1$) are very well separated in Fig.~\ref{fig:COFe}, top panel: a branch enriched by massive stars yielding [C/O]$\lessapprox -0.4$ and a wide range of [O/Fe]$\gtrapprox 1$, and the AGB enriched stars with [C/O]$\gtrapprox 0.5$ and [O/Fe]$\sim 0\hbox{--}2$ (where the Fe and most of the O comes from a small contribution from type~II SNe). 
 The bottom panel of the figure plots observed stars taken from the \saga\ database (filled circles show detections, crosses indicate upper limits, colour represents [Ba/H] as in the top panel). The data also show a large scatter in [C/O], with a trend of lower [Ba/H] stars having higher [O/Fe] which is also clearly seen in \eagle. We note that the two branches that stand-out in the simulation (top panel) are not so obvious in the data, however clearly there is an indication of the separation between type-s (bluer points, high barium) and type-no (redder points, low barium) stars.

 It is difficult to conclusively identify these two sub-classes of CEMP stars observationally in a diagram such as Fig.~\ref{fig:COFe}, because measuring the abundance of oxygen is more difficult than of other elements. The upper branch can be distinguished by the high s-process element abundance of such stars, a clear signature of the AGB origin of carbon in CEMP stars as we did in the previous section. The relative abundance pattern of $\alpha$-elements may provide a signature of the stars on the lower branch. This is explored in Fig.~\ref{fig:X}, where we compare results from \eagle\ in grey, to observed patterns of metal poor stars with [Fe/H]$<-2$ from the \saga\ database (coloured symbols); crosses indicate limits in cases of non-detection of one or more elements.

In the top left panel showing [C/O] versus [O/Fe], we have labelled the loci in which \eagle\ stars are predominantly enriched by AGB or by massive low-$Z~$ stars. Mixing these channels leads to the appearance of stars with [O/Fe]$\sim 0$, and [C/O]$\sim 0$.  The [C/O]$\sim 0$ value differs significantly from either AGB yields, [C/O]$\gtrapprox 1$, or yields from low-$~Z$ massive stars, [C/O]$\sim -0.9$. Although the upper AGB branch is well populated by observed stars, the presence of observed stars on the lower 	type~II branch may be less convincing.
	
However, a well documented feature of type~II SNe yields at low-$Z$ is the dramatic decrease in the scatter in [$\alpha$/Fe] along the sequence of increasing atomic number $A$ from O-Mg-Si to Ca. In the \cite{Woosley95} yields used in \eagle, this is clearly seen in the truncation of the grey region at large values of [$\alpha$/Fe] particularly for Si and Ca for low values of [C/$\alpha$]. The physical reason underlying this trend is that in the \lq onion\rq\ model of the SN precursor, the Ca shell lies close to the Fe core, whereas the Si, Mg and O shells, in that order, lie further away. If the central core is strongly bound, as is the case at low $Z$, the SN explosion may not be sufficiently energetic to expel deep stellar layers. In that case Ca and Fe should track each other much more tightly than O and Fe, say, which is indeed what the grey \eagle\ pattern shows. Note in particular the sharp reduction in the number of \eagle\ stars above [Si/Fe]$\sim 1$ or [Ca/Fe]$\sim 1$ at low [C/$\alpha$].

Interestingly, a sharp reduction in the scatter of [$\alpha$/Fe] for higher $A$ is also seen in the data. In addition, the number of observed stars dramatically decreases above [Mg/Fe]$=0.8$, [Si/Fe]$=1$ or [Ca/Fe]$>0.6$. Recall that in the \eagle\ simulation, AGB stars are the source of carbon in the stars with high [C/$\alpha$]. Therefore stars with low [C/$\alpha$] are enriched mainly by massive stars. Such stars are clearly present in the \saga\ data as well.

\section{Summary}
We have explored the origin of carbon enhanced metal poor (CEMP) stars in the \eagle\ cosmological hydrodynamical simulation \citep{Schaye15}, selecting galaxies by halo mass to be \lq Milky Way\rq-like. Data for Milky Way CEMP stars can be classified in a number of subclasses \citep{Beers05,Frebel15}, and we compared observed and simulated abundance patterns such as [C/O] versus [C/Fe] or [O/H]. Both simulation and data show a large increase in the scatter of [C/O] and an upturn in the median value of [C/O] at low [O/H]. 

The trends in the simulation are a consequence of two effects that relate to the nature of star formation in \eagle\ at high $z$: bursty star formation combined with poor metal mixing in low-mass galaxies. Stellar feedback powers strong outflows in \eagle\
	galaxies, particularly in those with maximum circular velocity $v_{\rm c,max}\leq 50$~km~s$^{-1}$ at $z>6$. The absence of gas following a star burst then prevents further star formation for sufficiently long times that, when eventually cosmological accretion replenishes the galaxy with mostly pristine gas, stars form but not before their natal gas is enriched by ejecta from asymptotic giant branch (AGB) stars. The simulation therefore yields two distinct classes of CEMP stars: AGB enriched stars, and stars enriched by massive stars whose descendant type~II SNe have low iron yields (which result from either massive low $Z$ progenitors, or $\approx 30$~M$_\odot$ more metal rich progenitors, in the models of \cite{Woosley95}).
	
The relatively large differences in the lifetimes of the progenitor stars that cause the enrichment ($\gtrapprox 300$~Myrs for the AGB stars, but $\lessapprox 10$~Myrs for a 20~M$_\odot$ star) then leads to the  prediction that the lowest [Fe/H] stars that are enriched first are of type CEMP-no, whereas CEMP-s stars form only at slightly higher [Fe/H]$\gtrapprox -3$, as is also observed \citep{Aoki07, Yong13}, as discussed by \cite{Frebel06}.  These classes  are also distinguished by their [C/H], (see Fig.~\ref{fig:CH} and \cite{Yoon16}), and by their [C/O] (see Fig.~\ref{fig:COFe}).

This scenario makes several testable predictions. A mixture of carbon enrichment by AGB and Fe-poor  massive stars is consistent with the large observed scatter in [C/O] at low [O/H], and is also evidence for poor (metal) mixing of the yields from the AGB and massive star enrichment channels. The observed abundances of $\alpha$-elements compared to carbon show similar trends with atomic number, $A$, as seen in \eagle; a dramatic decrease in the scatter of [$\alpha$/Fe] for Si and Ca, compared to O and Mg. In the simulations, this pattern is imprinted by low-$Z$ SNe type~II yields. The physical mechanism that underlies this is that the core region of massive low $Z$ stars is so strongly bound that its content is more difficult to eject by the SN explosion. This then also explains the correspondingly high [C/Fe] yields.
	
The abundance patterns of very low $Z$ stars are an imprint of the bursty nature of star formation at high $z$, and therefore may provide a handle on the nature of the galaxies that reionized the Universe. \cite{Sharma16a,Sharma16b} proposed a model whereby these high-$z$ starbursts drive outflows that clear channels through which ionising photons can escape, with binary stars potentially an important source of photons \citep{Stanway16}. This model may explain why the escape fraction of ionising photons increases dramatically with redshift, as is necessary if galaxies are the dominant 	sources of reionising photons \cite[e.g.][]{Haardt12, Khaire15}.
	
The bursty nature of these galaxies, combined with poor metal mixing, leaves signatures in the abundance patterns of the stars that formed at those early times, with CEMP-no stars forming predominantly during the gas rich burst phase, and CEMP-s stars forming during a  more quiescent phase. In other words, high-$z$ star formation determines the elemental abundances of low-$Z$ stars. This line of reasoning prompts us to speculate that CEMP stars were enriched by the stars that enabled galaxies to reionise the Universe.

\section*{Acknowledments}
We thank our \eagle\ colleagues (J. Schaye, M. Schaller, R. Crain and R. Bower) for allowing us to use the simulations. We also thank Max Pettini for insightful comments and Stefania Salvadorri for providing the data in her published paper, and for comments on an earlier draft. This work was supported by the Science and Technology Facilities Council [grant number ST/F001166/1], by the Interuniversity Attraction Poles Programme initiated by the Belgian Science Policy Office ([AP P7/08 CHARM]). We used the DiRAC Data Centric system at Durham University, operated by the Institute for Computational Cosmology on behalf of the STFC DiRAC HPC Facility (www.dirac.ac.uk). This equipment was funded by BIS National E-Infrastructure capital grant ST/K00042X/1, STFC capital grant ST/H008519/1, and STFC DiRAC is part of the National E-Infrastructure. The data used in the work is available through collaboration with the authors. M.~S. is supported by an STFC post-doctoral fellowship. R.~J.~C. is supported by a Royal Society University Research Fellowship.
\section*{Appendix}
\label{sect:appendix}
\begin{figure}
\includegraphics[width=\linewidth]{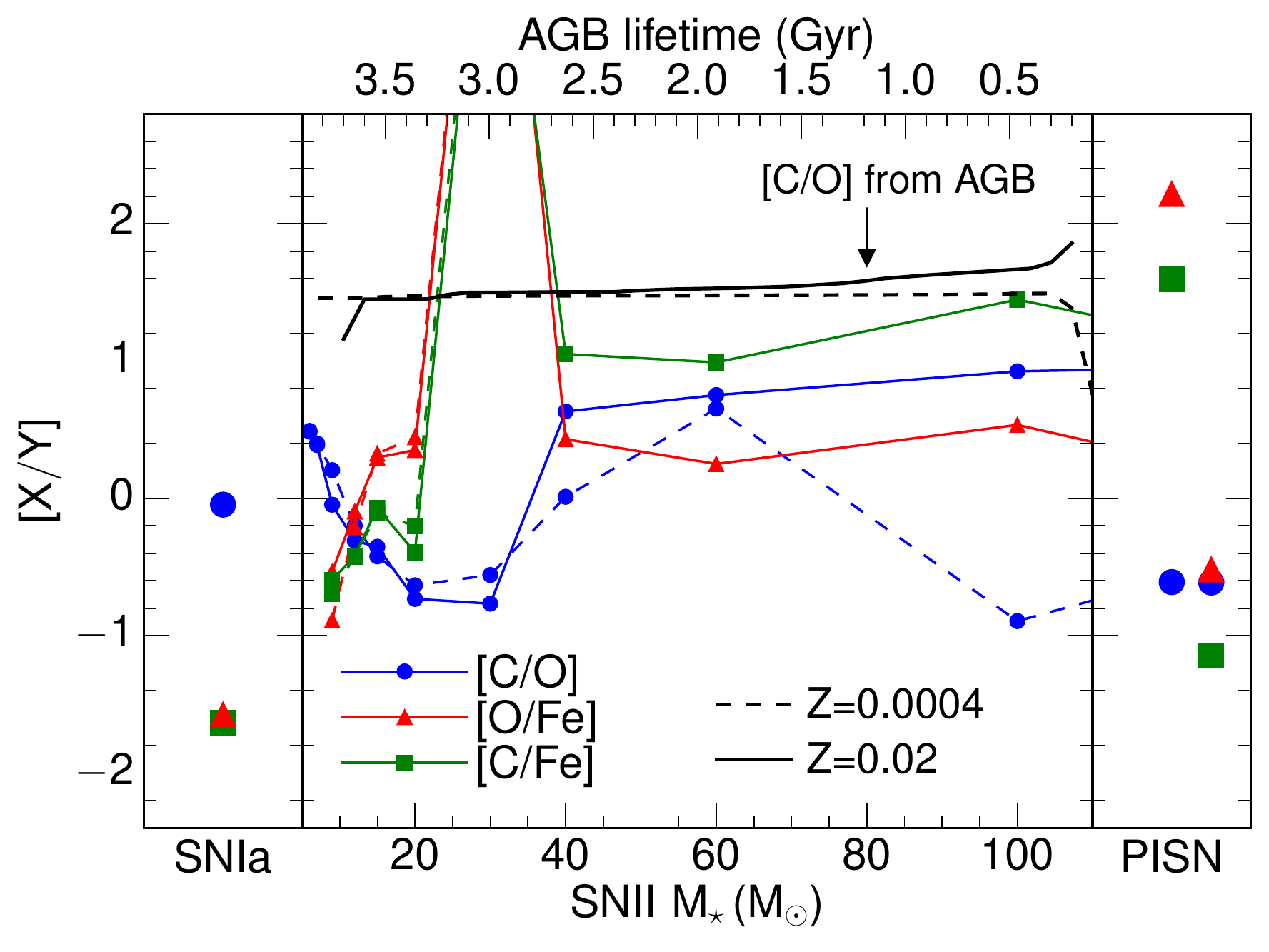}
\caption{Abundance ratios of stellar ejecta resulting from several nucleo-synthetic channels; see text for details.}

\label{fig:yield}
\end{figure}
The stellar evolution models and yields used in \eagle, implemented as described by \cite{Wiersma09}, are illustrated in Fig.~\ref{fig:yield}. Abundance ratios of [C/O], [O/Fe] and [C/Fe], are plotted in blue, red and green, respectively. Yields for type~Ia SNe, taken from model W3 of \cite{Thielemann03}, are plotted in the left inset.  Metallicity-dependent yields for type~II SN explosion from \cite{Woosley95} combined with the contribution from the progenitor star, are plotted as a function of the mass, $M_\star$, of the progenitor star in the central panel, for (solar abundance pattern) progenitors with $Z=0.02$ (solid lines), and $Z=0.0004$ (dashed lines). These models yield very low or no Fe for $M_\star\approx 30$~M$_\odot$ for either abundance. In addition, stars more massive than $20$~M$_\odot$ also do not produce iron for $Z=0.0004$. This is a consequence of the core of the SN precursor being so strongly bound that Fe is not ejected during the explosion (also called \lq fall back\rq). At (very) low $Z$ this occurs because of the absence of $^{12}$C to kick-start the CNO cycle when the proto-star heats up, consequently it contracts further to reach higher densities and temperatures. The central panel also shows the [C/O] yield of mass ejected by AGB stars as a function of lifetime (top $x$-axis, for solar abundance stars using the models of \cite{Marigo01} and \cite{Portinari98}. The right inset shows abundances for pair-instability SNe of progenitor mass $M_\star=150$~M$_\odot$ and $270$~M$_\odot$ (left and right set of points, respectively), taken from \cite{Heger02}.

\bibliographystyle{mnras}
\bibliography{ref_carbon}

\end{document}